\documentstyle[12pt]{article}
\makeatletter \@addtoreset{equation}{section}

\makeatother \textheight = 24truecm \textwidth = 16truecm \hoffset
= -1.3truecm \voffset = -2truecm

\begin{document}
\begin{center}
{\large {\bf Active Dwarf Galaxies as Circumnuclear Regions of
LSB-galaxies}} \vskip 0.5truecm Erastova L.K. \vskip 0.5truecm
 Byurakan
Astrophysical Observatory, Byurakan 378433, Aragatsotn District,
Armenia
\end{center}

\begin{abstract}
Some arguments are brought, that often active dwarf galaxies are the
circumnuclear regions of LSB galaxies in fact, rather than the ordinary
galaxies.
\end{abstract}

\section{Introduction.}

The important role of LSB (Low Surface Brightness) galaxies for
extragalactic researches came during the last two decades. They
are usually defined as galaxies with a blue central surface
brightness fainter than 21$^m $.65~\cite{A1}. However there are
the number definitions of LSB galaxies along their central and
average surface brightness. \\
A great number of LSB galaxies were
found on the plates of the new Palomar Observatory Sky Survey -
POSS2. What is more often the normal galaxies turned out to be as
LSB ones after discovery of LSB large halos, spiral structure and
other details on the deeper images.

\section{Hypothesis and observable facts.}

Suppose, that active dwarf galaxies with diameters no more than 5 kpc really
are not the normal formed galaxies. Probably they are the central parts, may
be buldges, of normal or even giant LSB galaxies. Suppose, that the
dimension of an average normal galaxy is about 20 kpc. In this case we can
see the next basic forms of flat active galaxies.
\\
First. Active region, having the size no more than 5 kpc, has the
weak halo up to 20 kpc in diameter. All morphological types of
flat systems with different types of activity, as Sey-type,
starbursts, BCGs and others may be constructed, changing the
surface brightness of this halo from values, which have not been
seen in optical light $\mu _0 > 26 ^m/arcsec$, when we can see
only ''naked'' active region, to the level of high surface
brightness.
\\
Second. It is the case, when some number of weak stellar associations there
are in the unseen halo.
\\
Third. The third case, when we can see only active nuclear part and one
great superassociation in the unseen LSB-halo.
\\
If the halo and/or stellar associations are very weak and they
can't be seen on our plates even with our largest optical
telescopes because for $\mu _0 \gg 26 ^m/arcsec$, then we can see
only buldge, circumnuclear part of really LSB-galaxy. Then we get
all types of active dwarf ''galaxies''.
\\
So we propose the next
tasks:

$\bullet$ The searching for weak halos or weak spiral structure
around active dwarf galaxies.

$\bullet$ The searching for weak blue stellar or semistellar
objects, probably stellar associations and superassociations near
active dwarf galaxies.
\\
Well, the examples of such systems had already been discovered.

$\bullet$ There are many active dwarf galaxies with huge unseen
halo. Only after obtaining deep images many active dwarf galaxies
show this halos. Therefore these active dwarf galaxies turned out
to be normal galaxies according to their dimensions. Such galaxy
often has very low mean surface brightness. There are Markarian
galaxies among them, Arakelian galaxies, which, by the way, were
discovered by their high surface brightness, KUG-Kiso ultraviolet
galaxies from Kiso survey, active galaxies from University of
Michigan (UM-galaxies)~\cite{A2}), Wasilewsky galaxies~\cite{A3}.
\\
In all these cases the galaxy changes its morphological type and active
region turn out to be either central part of galaxy or its
SA-superassociation.
\\
Let us consider the blue dwarf galaxies from the list~\cite{A4} or
~\cite{A5}. They are weak and often haven't any spectral data. But
it's not excluded, that blue or very blue very weak galaxies from
these lists are active dwarf galaxies or nuclear parts of normal
galaxies, but we haven't spectral or detailed images with good
resolution to say about active knots or hot spots in its.
\\
Therefore, going from normal galaxies to dwarfs we are going from full
galaxy to its central part region. So it's very difficult to describe the
morphology of dwarf galaxies, because they are circumnuclear regions of
galaxies, probably LSB-galaxies.
\\
Possibly, all above mentioned belongs to ordinary normal (not active
galaxies) (see, for example, UGC 9024 or NGC 628).
\\
Present examples show, that often really take place great LSB-halo
with active formation in it. It isn't already the dwarf galaxy.
But probably some of them are really active dwarf galaxies.
\\
Now at last there are intensive searches for galaxies, which are
the triggers of high star formation rate in BCDGs. Near some BCDGs
with low metallicity, such as, for example, Mkn 116=Izw 18, SBS
0335-052, HS 0822+3542 were discovered very weak blue dwarf
companions on the projected distances $< 20 kpc$ ( Mkn 116NW, SBS
0335-052E, SAO 0822+3545) (see~\cite{A6},~\cite{A7},~\cite{A8}).
It may be interpreted as the triggers of high star formation rate
in these galaxies. But may be they are the SA in these galaxies.
In this case these dwarf galaxies are not the ''dwarf'' ones, but
the huge LSB galaxies, where they are the circumnuclear parts, and
weak companions are the outer parts or stellar associations in
its.
\\
\section{Conclusions.}
1. Active dwarf galaxies are very similar by their structure,
morphology, dimensions, luminosities to the circumnuclear regions
of normal or even giant galaxies.
\\
2. There is the number of examples, when active dwarf galaxy turned out to
be active nuclear region of huge LSB-galaxy on the deeper plates.
\\
3. On this basis we propose, that active dwarf galaxies often are "naked"
nuclear region. Probably, they are the nuclear parts of LSB-galaxies, the
halos of which we can't observe even with the largest ground-based
telescopes.
\\
It is evident, that LSB and HSB galaxies are the extreme stages of galaxies.
Evidently there are all intermediate classes of galaxies. So, for example,
the host galaxies of Markarian galaxies often have lower surface brightness
and therefore their active nuclear regions have been easy discovered on the
survey plates.

\end{document}